# A CF$_3$I-based SDD Prototype for Spin-independent Dark Matter Searches

T. Morlat[a,1], M. Felizardo[a,b,c], F. Giuliani[a,2] TA Girard[a,*], G. Waysand[d],

R.F. Payne[e], H.S. Miley[e], A.R. Ramos[c,a], J.G. Marques[c,a], R.C. Martins[b],

D. Limagne[f]

[a] Centro de Física Nuclear, Universidade de Lisboa, 1649-003 Lisbon, Portugal
[b] Instituto de Telecomunicações, IST, Av. Rovisco Pais 1, 1049-001 Lisbon, Portugal
[c] Instituto Tecnológico e Nuclear, Estrada Nacional 10, 2686-953 Sacavém, Portugal
[d] Laboratoire Souterrain à Bas Bruit (Université de Nice-Sophia Antipolis), 84400 Rustrel-Pays d'Apt, France
[e] Pacific Northwest National Laboratory, Richland, WA 99352, USA
[f] INSP, Campus Boucicaut-140, 75015 Paris, France

(SIMPLE Collaboration)

## Abstract

The application of Superheated Droplet Detectors (SDDs) to dark matter searches has so far been confined to the light nuclei refrigerants $C_2ClF_5$ and $C_4F_{10}$ (SIMPLE and PICASSO, respectively), with a principle sensitivity to spin-dependent interactions. Given the competitive results of these devices, as a result of their intrinsic insensitivity to backgrounds, we have developed a prototype trifluoroiodomethane (CF$_3$I)-loaded SDD with increased sensitivity to spin-independent interactions as well. A low (0.102 kgd) exposure test operation of two high concentration, 1 liter devices is described, and the results compared with leading experiments in both spin-dependent and –independent sectors. Although competitive in both sectors when the difference in exposures is accounted for, a problem with fracturing of the detector gel must be addressed before significantly larger exposures can be envisioned.

1.     Introduction

The direct search for weakly interacting massive particle (WIMP) dark matter continues to be among the forefront endeavors of modern physics activity. Search experiments are traditionally classified as to whether spin-independent (SI) or spin-dependent (SD) according to which interaction channel the experiment is most sensitive, of which the first has generally attracted the most attention. The current status of the SI search for WIMPs is defined by a number of projects, including XENON [1], CDMS [2] and ZEPLIN [3], which as a result of their target nuclei spins also provide significant constraints on the WIMP-neutron sector of the SD phase space. Several new experiments with improved background discrimination capabilities and large active mass potential have been proposed with sensitivities intended to reach as low as $10^{-10}$ pb in the SI sector.

SIMPLE (Superheated Instrument for Massive ParticLe Experiments) [4] is one of only two experiments to search for evidence of WIMPs using freon-loaded superheated droplet detectors (SDDs) [5], and only a few with sensitivity to the WIMP-proton sector of the SD phase space, which is currently constrained by COUPP [6] and KIMS [7]. A SDD is a suspension of superheated liquid droplets (~ 30 μm radius) in a viscoelastic gel which undergo transitions to the gas phase upon energy deposition by incident radiation. Two conditions are required for the nucleation of the gas phase in the superheated liquid [8]: (i) the energy deposited must be greater than a thermodynamic minimum ($E_c$), and (ii) this energy must be deposited within a thermodynamically-defined minimum distance ($ar_c$) inside the droplet, where a is the nucleation parameter and $r_c$ = the thermodynamic critical bubble radius. The two conditions together require energy depositions of order ~150 keV/μm for a bubble nucleation, rendering the SDD effectively insensitive to the majority of traditional detector backgrounds which complicate more conventional dark matter search detectors (including γ's, β's and cosmic muons). This insensitivity is not trivial: given the ~ $10^7$ evt/kgd environmental γ rate observed in an unshielded 1 kg Ge detector [9], the blindness to γ's is equivalent to an intrinsic rejection factor several orders of magnitude larger than the bolometer experiments with particle discrimination. Both superheated liquid projects have demonstrated a potential to achieve competitive results with significantly reduced measurement exposures.

The refrigerants employed by SIMPLE and PICASSO are $C_2ClF_5$ and $C_4F_{10}$, respectively. Because of their fluorine content and fluorine's high spin sensitivity, and their otherwise light nuclei content, these have generally been considered exclusively SD search activities. Given their performance, together with the relative inexpensiveness of the technique, the question naturally arises as to whether or not SDDs might have a similar impact in the spin-independent sector. Since the SI cross section generally scales with the squares of both the mass number and the WIMP-nucleus reduced mass, exploring this sector of WIMP interactions suggests a SDD composition with nuclei of a significantly higher mass number [9]. The difficulty here is that device fabrications generally proceed on the basis of density-matching the refrigerant with the gel, and that heavier refrigerants are generally of higher density. The traditional addition of heavy salts such as CsCl to raise the gel density is here discouraged, given that this adds radioactive contaminants which must be later removed chemically. Thus, although several readily available "heavy" refrigerants exist, such as $CF_3Br$, $CF_3I$ and $XeF_4$, the problem of fabricating a homogeneous dispersion of the refrigerant, without introducing additional radio-contaminants, has discouraged their development. This problem has been circumvented by COUPP using a $CF_3I$ bubble chamber technique (which brings its own problems); its recent first result [6] has yielded restrictive constraints on the spin-dependent sector comparable with the significantly larger exposure KIMS.

Recently, we described the potential impact of an SDD search based on $CF_3I$, following our initial fabrication of a small volume (150 ml), high concentration (1-3 %) prototype using viscosity rather than density matching [10]. We here report the fabrication details (Sec. 2) and testing, and a low exposure underground pilot test (Sec. 3) of two 1 liter devices. The results are compared with the previsions of Ref. [10] and with the recent COUPP result, and their implications assessed in Sec. 4. The conclusions identify several problems associated with this particular detector which must be addressed prior to its search application.

## 2. Detector Fabrication

The variation of the $CF_3I$ density with temperature is shown in Fig. 1, relative to that of the standard SIMPLE $C_2ClF_5$. The significant difference in gel and refrigerant densities (1.3 and 2 gcm$^{-3}$ at 40ºC, respectively) would generally cause the refrigerant droplets to sink during fabrication, resulting in an inhomogeneous distribution of droplet sizes located near the detector bottom.

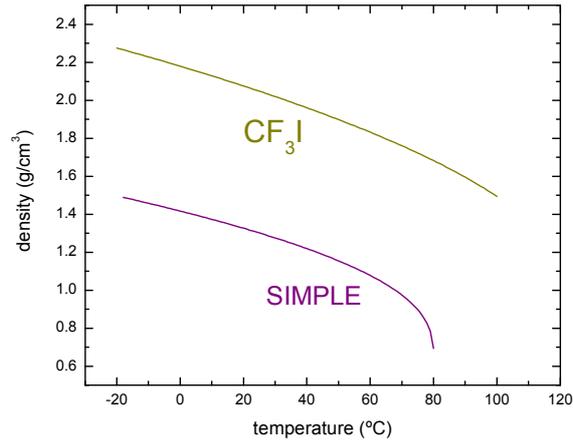

Fig. 1 : Variation of the CF$_3$I refrigerant density with temperature, relative to the standard SIMPLE refrigerant C$_2$ClF$_5$.

An alternative approach, at least in principle, is to match in viscosity rather than density. An estimate of the minimum viscosity (η) required to trap the droplets during the fabrication process is obtained by equating the viscous ($6\pi r \eta \frac{D}{t}$) and gravity-Archimede's ($\frac{4}{3}\pi[\rho_b - \rho_0]r^3 g$) forces, where r is the average droplet radius, D is the height of the gel, t is the time for a droplet to fall a distance D, and ρ$_b$ (ρ$_0$) is the CF$_3$I (gel) density. This gives

$$\eta = 2r^2 gt \frac{\rho_b - \rho_0}{9D} \quad . \tag{1}$$

For t = 1 hour (the time required for the setting of the gel during cooling), ρ$_b$ (ρ$_0$) = 2 x 10$^3$ kg/m$^3$ (1.3 x 10$^3$ kg/m$^3$), r = 35 x 10$^{-6}$ m and D = 5 x 10$^{-2}$ m, η = 0.13 kg/m/s.

Given this viscosity estimate, numerous small volume (150 ml) prototype devices were made and examined, varying the basic fabrication recipe and protocol of the previous C$_2$ClF$_5$ fabrications [4]; the viscosity variations were effected with agarose, blending a small quantity with glycerin at a temperature of 90°C to break the agarose chains prior to its addition to the gel. The gel itself was formed by combining powdered gelatin and bi-distilled water with slow agitation to homogenize the solution. Separately, PVP was added to bi-distilled water, and agitated at 60°C. Both gel ingredients contained pre-eluted ion-exchange resins for actinide removal, which were removed by filtering [11] after blending in a detector bottle by agitation. The glycerin + agarose solution, after its own filtration, was then slowly added to the gel. Following outgassing and foam aspiration, the solution was left overnight at 42°C with slow agitation to prevent air bubble formation. The final gel matrix recipe, which later produced a uniform and homogeneous distribution of droplets, had a measured $\eta = 0.17$ kg/m/s, as well as an increased temperature at which the transition from solution to gel (sol-gel transition) occurs.

SDD fabrication occurs via the phase diagram of Fig. 2. The detector bottle was removed to a hotplate within a hyperbaric chamber, and the pressure raised to just beyond the vapor pressure at 42°C.

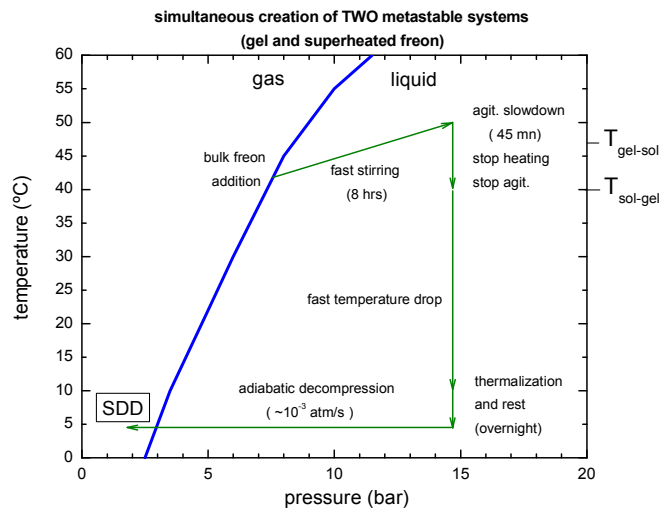

Fig. 2 : Phase diagram of the SDD fabrication procedure, as described in text; note the hysteresis in the sol-gel transition.

After thermalization, the agitation was stopped and the $CF_3I$ storage bottle opened. The pressure difference between bottle and chamber moved the $CF_3I$ into the chamber, with the flowline immersed in ice to condense and distill it at the same time. The $CF_3I$ was also forced through a micro-syringe filter of 0.2 μm [12] before falling into the gel.

Once the $CF_3I$ was injected, the pressure was quickly raised to 15 bar to prevent the droplets from rising to the surface, and a rapid agitation initiated to shear big droplets; simultaneously, the temperature was raised to 50°C to create a temperature gradient inside the matrix and permit dispersion of the droplets. After 20 minutes, the temperature was slightly reduced for 1 hr (with pressure and agitation unchanged). The $CF_3I$, in liquid state, was divided into smaller droplets by the continued agitation. Finally, the heating was stopped: the temperature decreased until the sol-gel transition was crossed, during which the stirring was reduced and finally stopped. The droplet suspension was quickly cooled to 10°C and left to set for 40 minutes, then cooled to 5ºC where it was maintained for ~ 15 hours. The pressure was then slowly reduced to atmospheric pressure, and the detector removed to cold storage. The process resulted in approximately uniform and homogeneous (40±15 µm diameter) droplet distributions, as determined by optical microscopy. Longer fractionating times give narrower distributions of smaller diameters; shorter, broader distributions of larger diameters.

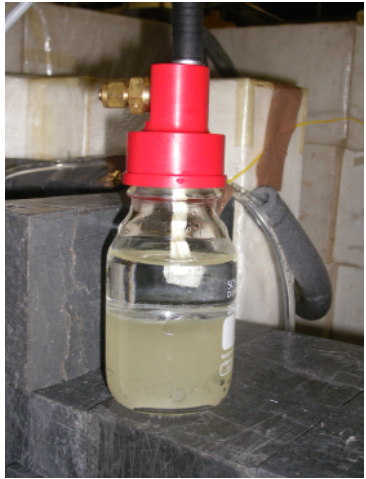

Fig. 3 : Completed 150 ml $CF_3I$ detector prototype, with new capping. Note the encapsulated microphone in the glycerin layer.

### 3.    Prototype Test Measurements

#### 3.1    Irradiations

Irradiations of the small volume device prototypes by $^{60}$Co verified the device insensitivity to γ's below ~ 45ºC (the temperature at which the gel melts), consistent with the general response of SDDs. Irradiations with a filtered neutron beam from the thermal column of the RPI [13] demonstrated sensitivity to reactor neutron irradiations via the induced recoils of fluorine, carbon and iodine. Fig. 4(a) displays the results of a 144 keV neutron irradiation of a device at 1 bar, with the rapid rate increase ~ 40ºC consistent with the iodine sensitivity onset observed in the temperature variation of the threshold incident neutron energies ($E_0^A$) in Fig. 4(b). These were calculated from the threshold recoil energy ($E_{thr}^A$) via $E_0^A = [(1+A)^2/4A]E_{thr}^A$, and are also shown in Fig. 4(a) above the corresponding temperature. The reduced superheat factor [14], $s = [(T - T_b)/(T_c - T_b)]$ with $T_c$, $T_b$ the critical and boiling temperatures of the refrigerant at a given pressure, is displayed below. The expected signal from fluorine and carbon at 20ºC is masked by the iodine response to a broad, higher energy neutron component of the filtered beam, as identified in Ref. [15].

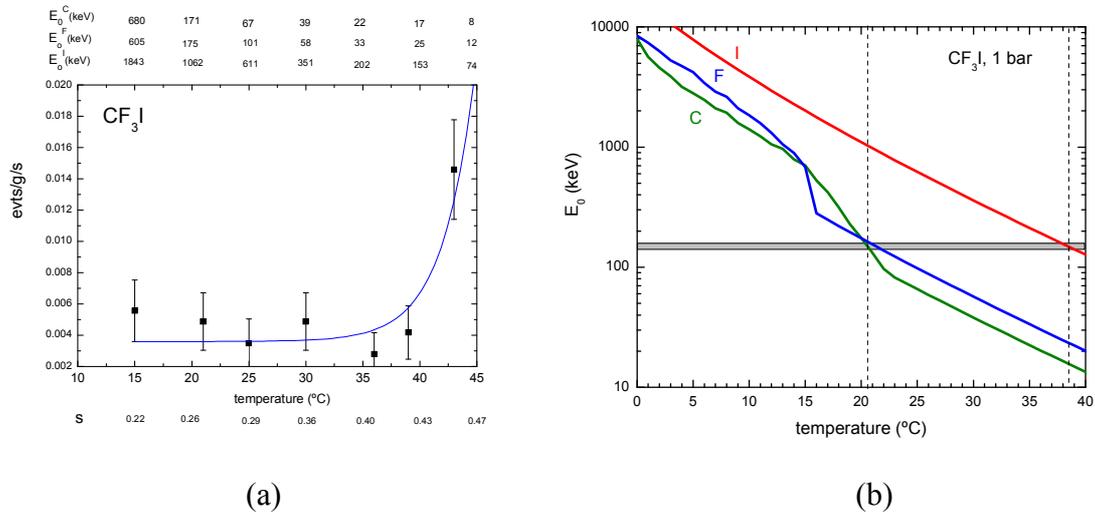

|         | (a) | (b) |
|---------|-----|-----|

Fig. 4 : (a) 144 keV filtered neutron irradiation of a $CF_3I$ prototype below the gel melting at 45ºC; the line represents an exponential fit to the data. (b) Variation of the threshold incident neutron energies ($E_0^A$) with temperature, with the shaded central band indicating 144±20 keV incident neutron energy; the dotted lines indicate the intersections of the two.

Unlike previous detectors made with $C_2ClF_5$, the prototypes began to significantly fracture within several hours of fabrication. The fracturing is inhibited by overpressuring the devices, but not eliminated. Tests with a SDD made by dissolving the refrigerant inside the gel produced cracks within 24 hrs, indicating the fracturing to occur because of a high solubility of $CF_3I$ gas inside the gel, despite the manufacturer's claims of insolubility (our studies suggest ~ 0.5 g/kgH$_2$O/bar). Although this phenomenon occurs with or without bubble nucleation, because the $CF_3I$ gas inside the gel occupies any microscopic $N_2$ gas pockets formed during the fractionating stage of the suspension fabrication, it is aggravated by nucleations arising from the ambient background radiations of the surface, reactor fabrication site.

Optical microscopy further identified a significant presence of clathrate hydrates at low temperature, which tend to "glitter" because of the ice crystal cages when observed through a microscope. Clathrate hydrates are crystalline, water-based solids physically resembling ice, in which small non-polar molecules (typically gases) are trapped inside

"cages" of hydrogen bonded water molecules. Their formation and decomposition are first order phase transitions, not chemical reactions; they are not chemical compounds as the trapped molecules are not bonded to the lattice. Their presence implies that the device cannot be stored in a stable liquid phase at temperatures below 0ºC (as previously done with the SIMPLE devices which were transported from their Paris fabrication site) because clathrate hydrates provoke spontaneous nucleation locally on the droplet surfaces in warming to room temperature, effectively destroying the device.

### 3.2 Underground test measurement

Assuming a longer lifetime to be obtained in a shielded location, a further test was effected with 1 liter device prototypes fabricated in a recently-installed 210 mwe deep clean room facility of the Laboratoire Souterrain à Bas Bruit (LSBB) [16] during the facility's commissioning. The ~ 1 km proximity to the 1500 mwe deep measurement site also reduced difficulties with device damage during transportation.

Two detectors were manufactured according to the recipe and protocol of Sec. 2, one containing 6.1 g $CF_3I$ and the other, ~ 19.5 g (primarily intended for fracture studies). Both exhibited a small (~1%) asymmetry in the top-to-bottom droplet density under microscopic examination. The devices were capped using a new mechanical construction designed to suppress the previous problems with pressure microleaks [4,17] through the plastic SDD caps of the submerged devices, which yielded signals indistinguishable from bubble nucleation events. While the majority of these were eliminated by coincidence between the detector transducers, the new cap construction has shown no microleaks in tests of up to three weeks submerged operation, yielding a 90% C.L. upper limit of ~ 0.11 microleaks/detector/day, versus the previous 0.5-1 microleaks/detector/day.

As in previous SIMPLE measurements [4], both devices were pressurized to 2 bar following fabrication to reduce background sensitivity, transferred to the 1500 mwe level of the LSBB, and centrally installed inside a 700 liter water bath, which was surrounded by three layers of sound and thermal insulation, and rested on a dual vibration absorber. The 19.5 g device manifested three large fractures at the time of installation; the 6.1 g device, in contrast, was pristine.

Each detector was instrumented with a single piezoelectric transducer (PKM-13EPY-4002-Bo), encased in a latex sheath and immersed in a glycerin layer at the top of the device, which recorded the acoustic shock wave accompanying the bubble nucleation. Rear-end electronics amplified the transducer signal by a factor of $10^5$; their time-tagged waveform traces were recorded in a Matlab platform. The transducer electronics was a hybrid of that used in earlier SIMPLE instrumentation [4, 18], having the second stage of amplification and the band-pass filter removed, which reduced the overall noise level by ~$10^2$ and provided a discrimination capability between true nucleation events and microleaks [18] via their fast Fourier transforms, as shown in Fig. 5.

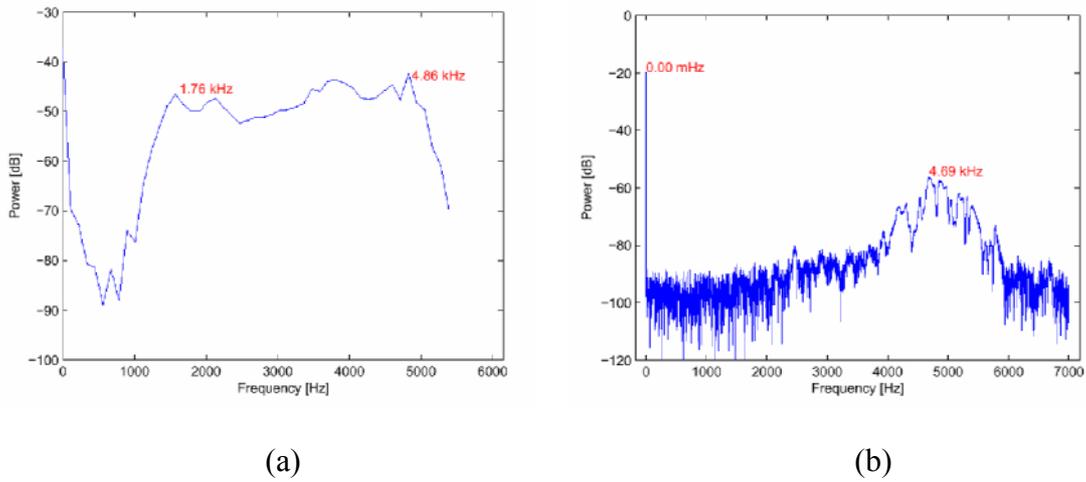

(a)  (b)

Fig. 5 : Typical fast Fourier transforms of microleak (a) and true bubble nucleation (b) events.

The improved electronics yielded a noise level of 1 mV, in contrast to the > 13 mV levels recorded above ground and ~ 100 mV of the previous instrumentation. The use of telecommunications-grade cabling yielded no pickup resulting from ambient acoustic noise, cable motion or activity in the site, even when exaggerated, as experienced in previous measurements.

The detectors were continuously monitored over a 5 day period at an ambient bath temperature of $16^{o}$C. The test was briefly interrupted at various junctures for visual inspection of the detectors and performance checks to verify that both detectors were functioning.

In the first analysis stage, the detector responses were inspected with respect to raw signal rate, threshold behavior, and pressure evolution over the measurement period. A data set corresponding to 4 days was extracted from the device records. Events coincident in the two devices were next rejected. The result was then subjected to a pulse shape analysis [18], in comparison with a template of true bubble nucleation events. Figure 6 shows a typical waveform from the experiment (a), in comparison with a true bubble nucleation event (b) thermally-stimulated offline, indicating amplitude differences of > 10 and a significantly different pulse structure. This yielded 11 candidate events in the small device, and 15 in the larger. The number of recorded events equaled the number of new bubbles observed during the detector removal from the water bath following the test, consistent with a 100% detection efficiency observed in previous SIMPLE device calibrations.

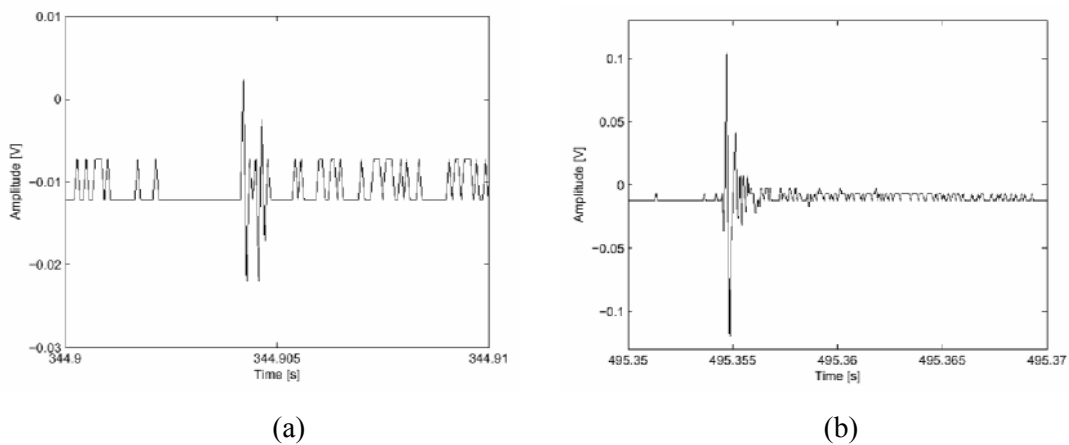

(a)                                        (b)

Fig. 6 : Typical waveform of a generic measured signal after first and second
stage filtering (a) and a true bubble nucleation event (b);
note the difference in amplitude and structure.

Finally, a fast Fourier transform analysis of the surviving candidates was made, a true nucleation event of which possesses a well-defined frequency response, with a time span of a few milliseconds and a primary harmonic at ~ 4.8 kHz [18] as seen in Fig. 5(b). This filtering stage yielded no candidate events in either detector.

**4.     Assessment and Discussion**

*4.1     Background estimates*

Analysis of the background contributions was extensively considered in Ref. [10]. In summary, the response of SDDs, of both low and high concentrations, to γ's, β's and cosmic-ray muons is similar [14,19], with a threshold sensitivity to these backgrounds occurring for s ≥ 0.5. The $CF_3I$ operation at 16°C and 2 bar corresponds to s ~ 0.16, sufficiently below threshold for this contribution to be neglected.

At 1500 mwe, the ambient neutron flux is primarily a fission spectrum from the rock, estimated at well-below $4 \times 10^{-5}$ n/cm²s [20]. The 50 cm surrounding water bath additionally acts as a neutron moderator, as also the gel, further reducing any ambient neutron flux by at least four orders of magnitude. The response of low concentration SDDs to various neutron fields has been studied extensively [14,21,22]; the high concentration SDD response to neutrons has been investigated using sources of Am/Be, $^{252}$Cf [17,19] and monochromatic low energy neutron beams [15,19]. This yields an estimate of $< 3 \times 10^{-2}$ neutron events for the 4 day measurement.

The metastability of a superheated liquid, as described by the homogeneous nucleation theory [23], gives a stability limit of the liquid phase at approximately 90% of the critical temperature for organic liquids at atmospheric pressure; at 40°C and 2 bar, the theoretical probability is $10^{-1800}$ nucleations/kgd and decreases with decreasing temperature. At s = 0.16, given the high purity and the smooth droplet/gel interfaces, the inhomogeneous contribution to spontaneous droplet nucleation is also entirely negligible.

The most probable primary internal background is α and α-recoils induced by radio-contaminants in the gel and detector glass. The current SIMPLE gel ingredients used in the $CF_3I$ prototype, all biologically-clean food products are purified using pre-eluted ion-exchanging resins specifically suited to actinide removal; the freon is single distilled; the water, double distilled. The presence of U/Th contaminations in the gel, measured at ≤ 0.1 ppb by low-level α and γ spectroscopy [24-26] of various samples of the production gel, yields an overall α-background level of < 0.5 evts/kg freon/d. The α response of SDDs has been studied extensively [17,19]. The SRIM-simulated dE/dx for α's in $CF_3I$ has a Bragg peak at 700 keV and ~ 193 keV/μm, which sets the temperature threshold for direct α detection with $CF_3I$; below this threshold, α's can only be detected through α-induced nuclear recoils. Radon contamination is low because of the surrounding 50 cm of circulating water bath which reduces the entry of atmospheric radon (28 Bq/m³): radon solubility in water is 230 cm³/kg at 20°C and decreases with temperature increase; with a diffusion length of 2.2 cm, the radon concentration is reduced by a factor 10 at 7 cm below the water surface. It is also low because of the short (< 0.7 mm) radon diffusion lengths of the SDD construction materials (glass, metal), with the measured radon contamination of the glass at a level similar to that of the gel. The N2 overpressuring of the device inhibits the advective influx of Rn-bearing water through the device capping, as well as diffusion of Rn from the walls of the glass container into the gel (via stiffening of the gel). Given the short time of this measurement, and insensitivity of the SDD to the β's of the radon decay chains, the overall α contribution is < 0.1 evts.

The 90% C.L. upper limit of ~ 0.11 microleaks/detector/day with the new detector capping corresponds to ~ 0.44 events for the measurement. In short, the estimated total background contributions, including microleaks, give ≤ 1 event for the measurement. The most probable origin of the rejected signals is the slow increase of fracture lines during the measurement: the expansion of the refrigerant into microscopic gas pocket cavities causes a visible and audible crack (as observed in the SDD tests with the refrigerant fully dissolved in the gel mentioned above).

*4.2    Recoil sensitivity*

The threshold recoil energy ($E_{thr}^A$) dependence on temperature of the target nuclides is shown in Fig. 7, calculated using thermodynamic parameters taken from Ref. [27], recoiling ion stopping powers precalculated with SRIM 2003 [28]; although the nucleation parameter a = 4 [29] was used in Ref. [10], we here use a = 6 [6].

As discussed in Ref. 10, the bubble nucleation efficiency of an ion of mass number A recoiling with energy E is given by the superheat factor $S_A(E) = 1-E_{thr}^A/\{E\}$ [22]. $E_{thr}^A$ can be set as low as 6.5 keV before onset of the gel melting at 45°C; the stopping power is ≥ 100 keV/μm for temperatures up to ~ 5°C above the gel melting point. As evident, at temperatures above ~ 22°C (a = 6), the $E_{thr}^A$ and $S_A$ are the same for fluorine, carbon, and iodine; whenever all recoiling ions stop within a pressure- and temperature-dependent critical distance, $E_{thr}^A$ and $S_A(E)$ do not depend on A. The reason that the $E_{thr}^A$ of iodine, fluorine and carbon ions in Fig. 7 do not coincide for all temperatures is that while in the range $E_{thr}^I = E_c$, below ~ 19°C a fluorine ion above $E_c$ and below $E_{thr}^F$ has insufficient stopping power to trigger a nucleation. More generally, a particle above $E_c$ but below the stopping power threshold cannot directly produce a bubble nucleation, and can only be detected indirectly, with lower efficiency, through a secondary recoiling ion.

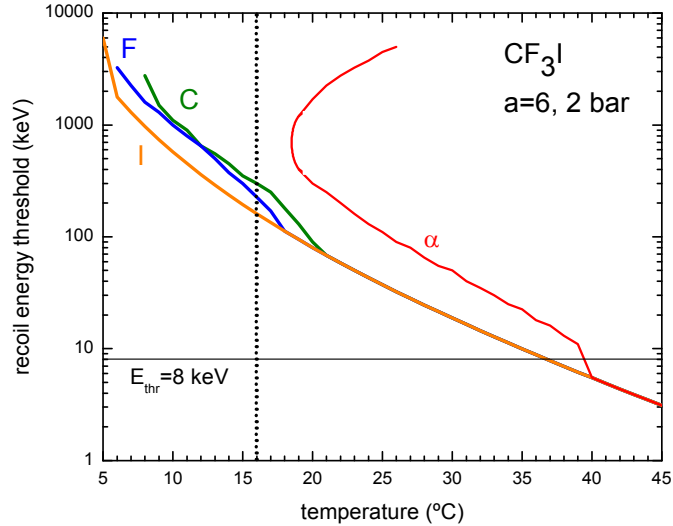

Fig. 7 : Variation of the threshold recoil energy with temperature for the three CF$_3$I constituents at 2 bar, obtained using a nucleation parameter of 6. The curve labeled "α" indicates the thresholds between which the device is sensitive to α's.

Figure 7 also shows an additional curve labeled "α", defining the threshold below which the SDD is insensitive to α's for a = 6. In Ref. [10], the α-sensitivity threshold appearing with the threshold curves (obtained with a = 4) was in fact calculated using a = $4.3(\rho_V/\rho_l)^{1/3}$ with $\rho_V$ ($\rho_l$) the refrigerant vapor (liquid) density [30], and represented only the minimum temperature for α sensitivity onset; here, the full threshold curve from the Bragg peak is shown. While an 8 keV threshold recoil energy remains at a temperature of 37ºC in both cases, a = 6 yields an α-sensitivity threshold of ≥ 20 keV; at 16ºC, there is no α-sensitivity.

Given the above considerations, the upper limit on the WIMP rate at 90% C.L., obtained from a 90% probability estimate for a Poisson distribution, is then (ln10)/0.102 = 22.6 evt/kgd. From Fig. 7, the minimum threshold recoil energy at 16ºC and 2 bar is 175 keV from the iodine curve; the fluorine and carbon were insensitive to WIMPs since the maximum WIMP-induced recoil energy (assuming a standard isothermal halo) is ~ 180 and 280 keV, respectively, both below the respective threshold recoil energies at 16ºC and 2 bar operating pressure.

### 4.3 Spin-dependent sector

The impact of the test on the SD phase space is shown in Fig. 8(a), calculated within a model-independent formulation [31,32] using the standard isothermal halo and WIMP scattering rate [33] with zero momentum transfer spin-dependent cross section $\sigma_{SD}$ for elastic scattering:

$$\sigma_{SD} = \frac{32}{\pi} G_F^2 \mu^2 \left[ a_p \langle S_p \rangle + a_n \langle S_n \rangle \right]^2 \frac{J+1}{J} \quad , \quad (2)$$

where $\langle S_{p,n} \rangle$ are the expectation values of the proton (neutron) group's spin, $G_F$ is the Fermi coupling constant, and J is the total nuclear spin. The form factors of Ressell and Dean [34] have been used for iodine, and those of Lewin and Smith [33] for fluorine and carbon, In this formulation, the region excluded by an experiment lies outside the indicated band, and the allowed region (shaded) is defined by the intersection of the various bands. We use the spin values of Strottman et. al. [35]; use of the Divari et. al. values [36] would rotate the ellipse about the origin to a more horizontal position, and make it thinner. Masses above or below this choice yield slightly increased limits.

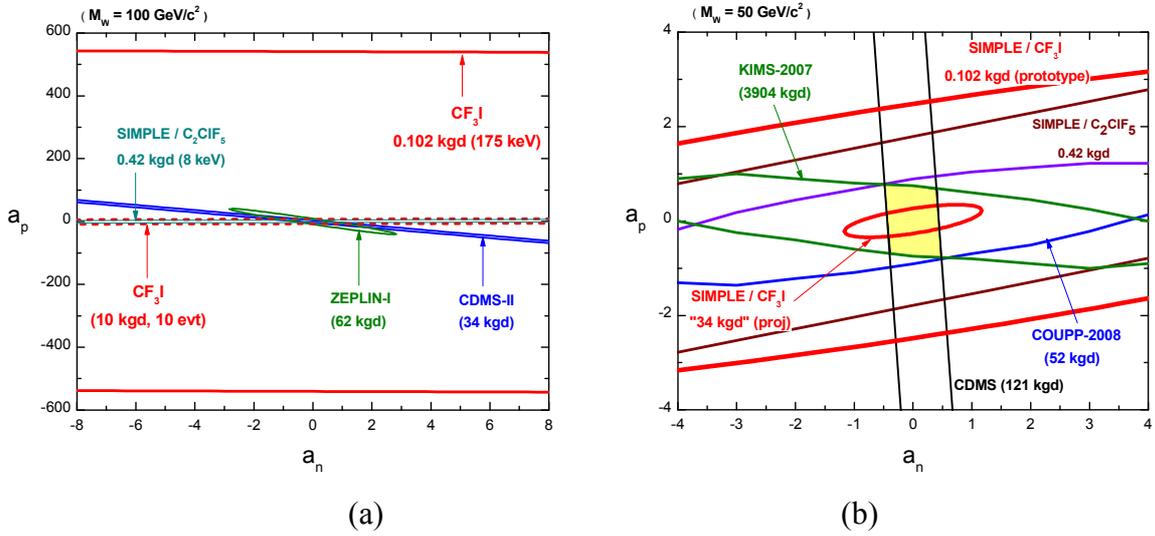

Fig. 8 : (a) $a_p$-$a_n$ for the SIMPLE/CF$_3$I at $M_W$ = 100 GeV/c$^2$ , indicating $|a_p| < 600$; (b) at $M_W$ = 50 GeV/c$^2$ with the same test result at 37°C (8 keV threshold) operation, together with benchmark experiment results. Also shown, a 34 kgd projection (dotted) assuming an undiscriminated 1 evt/kgd background. The region permitted by each experiment is the area inside the respective contour, with the allowed area at 50 GeV/c$^2$ in yellow.

As evident, there is no impact, owing to the spectator role of the fluorine at 16ºC (see Fig. 7), and small spin values of the sensitive iodine. Increasing the temperature would provide an immediate improvement in the sensitivity: at 37ºC all target nuclei would be responsive (see Fig. 7). The temperature remains below the sensitivity threshold for muon/electron, and γ backgrounds discussed above, as well as the onset of gel melting at 45ºC and threshold for spontaneous nucleation. As seen in Fig. 7, at 37ºC the SDD is fully sensitive to alpha's > 40 keV in recoil energy; given the measured radiocontamination levels of the purified gel and glass, the simulated alpha rate is < 0.7 evt/kg freon/d, or < 1 x$10^{-2}$ evts/detector/d. Although this is a problem for long exposure measurements, the discrimination of such events via amplitude analysis of their acoustic signal has recently been demonstrated [37]. The increase of sensitivity with temperature increase for neutron backgrounds below ~600 keV in incident energy (see Fig. 4(a)) would be offset by the 50 cm of water shielding. Figure 8(b) indicates this improvement, assuming the same null result at 37ºC, together with the results from CDMS, KIMS and COUPP.

In general, for the same background rates, the results scale as $\sqrt[4]{exposure}$, suggesting the same 0.42 kgd exposure to reach the contour of the $C_2ClF_5$ result. Fig. 8(b) also includes a projection for a 34 kgd exposure with 8 keV recoil threshold assuming a 1 evt/kgd undiscriminated background, as described in detail in Ref. [10]. This is consistent with scaling from the 0.102 kgd contour, suggesting a significant impact on the spin-dependent phase space with relatively small exposures. Surprisingly, the 34 kgd contour lies fully within that of the COUPP result, indicating the latter to have a significant higher level of un-discriminated background (likely due to the current α-backgrounds as discussed in Ref. 6).

### 4.4   Spin-independent sector

The impact of the test on the SI sector is shown in Fig. 9 in comparison with results from other leading search efforts. The contour is calculated following the standard isothermal halo and WIMP elastic scattering rate of Ref. [33] using a Helm form factor, as described in Ref. [10]. The figure indicates a contour minimum of ~ 0.1 pb at ~ 200 GeV/$c^2$.

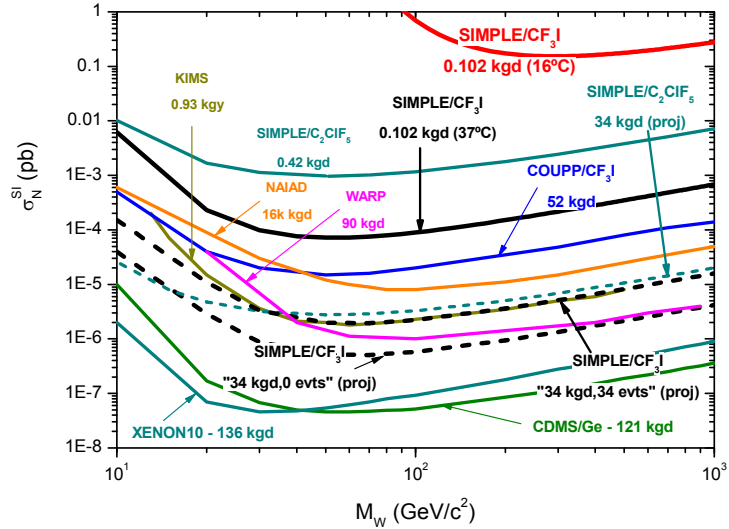

Fig. 9 : Spin-independent contour for SIMPLE/CF$_3$I (solid), together with several of the leading spin-independent search results. The dashed line indicates the contour which would have been obtained at 37°C, assuming the same exposure and null result. The projected 34 kgd SIMPLE/CF$_3$I exposures are shown (dotted) for comparison, for both a 1 evt/kgd background level ("34 evts") and full background discrimination ("0 evts") [10]. Also shown is the impact of the 52 kgd COUPP result [38].

The region below 100 GeV/c$^2$ was essentially inaccessible in this test, owing to the high recoil threshold associated with the operating temperature. As in the SD sector, increasing the temperature would provide an immediate improvement in the sensitivity (1.6x10$^{-4}$ pb, with a contour minimum of 6.8x10$^{-5}$ pb at 50 GeV/c$^2$), as indicated by the dotted line which assumes the same null result at 37°C.

The impact of the iodine presence and null result is seen in the nearly factor 10 improvement in sensitivity of the 37°C contour projection of this result and that of the previous C$_2$ClF$_5$ result with a factor 4 more exposure and background of 30 evt/kgd. At 300-500 GeV/c$^2$, even the current result is only a factor 10-30 less sensitive than the previous, consistent with the exposure difference and higher recoil threshold [39]. To reach the contours of either CDMS or XENON10 at 50 GeV/c$^2$ would require ~170 kgd exposures, respectively, roughly equivalent to those reported.

The current COUPP impact in the SI sector is also shown, which lies less than an order of magnitude below that of the 37°C projection despite the factor 500 difference in exposures. This again indicates an un-discriminated background rate significantly above 1 evt/kgd. The greater sensitivity of COUPP at low $M_W$ further indicates a recoil threshold operation below 8 keV.

## 5. Conclusions

Prototype high concentration, large volume $CF_3I$-based SDDs have been fabricated, and tested at 1500 mwe for 5 days with an $E_{thr}$ = 175 keV. Improved acoustic instrumentation provides no evidence of microleaks, and yields no events consistent with the characterization of a true bubble nucleation event.

This test result was not intended to contribute to the current restrictions of either SI or SD WIMP existence, but rather provide an experimental basis for the assessment of device employment complementing Ref. [10]. The nucleation parameter for $CF_3I$ is critical to both the conduct of the measurement and the analysis of its results, given an ~8°C shift to higher temperature in the threshold curves of Fig. 7 with a =4. The a = $4.3(\rho_V/\rho_l)^{1/3}$ has been shown to be in agreement with experiment for $C_2ClF_5$ and $CCl_2F_2$, but is not confirmed with $C_4F_{10}$. Although COUPP reports an experimental a = 6, this lacks any thermodynamic dependence of theory, and appears to vary with operating pressure [40]: a new measurement of this parameter is essential towards clarifying the situation. Clearly however, operating the devices at temperatures near 37°C, resulting in an 8 keV recoil threshold, would have provided almost a $10^3$ greater sensitivity in the SI, and $10^2$ improvement in SD. The results show that a larger exposure measurement with such a device would be capable of providing competitive restrictions in both sectors of the search for WIMP dark matter, with relatively lower exposures and less expense.

Larger exposures however appear problematic because of the deterioration of the current device. Unlike in the $C_2ClF_5$ fabrications, the results indicate a large fraction (~50%) of the refrigerant to dissolve into the gel due to its high solubility in the weak hydrogen bond gel matrix. This leads to a fracturing of the gel, with or without bubble nucleation, and performance degradation. This fracturing is slowed by overpressuring the devices, but not eliminated. There is also a significant presence of clathrate hydrates at low temperature, implying that the device cannot be stored below 0°C.

The problem of the clathrate hydrates can be circumvented by fabricating the devices in close proximity to the measurement site. Underground fabrication and operation also reduce the fracturing rate, and newly developed SDD instrumentation based on a true electret microphone has shown a capability to discriminate not only microleaks, but also events associated with device fracturing [41]. Nevertheless, the problem of fracturing must be addressed in order to achieve the larger exposures required of a full-scale search. This requires an improved understanding of the involved chemistry and development of new techniques, to include the possible use of gelifying agents not requiring water as a solvent or the use of ingredients to inhibit the diffusion of the dissolved gas. The high solubility of the $CF_3I$ in the gel implies the elimination of the water, which in turn suggests possibly shifting to organic gels if the present radio-purity of the current fabrications can be maintained. Also under investigation are new constructs in the detector fabrication which would eliminate the fracturing entirely.

## Acknowledgments

We thank M. Auguste, D. Boyer, and A. Cavaillou of the Laboratoire Souterrain à Bas Bruit (Université de Nice-Sophia Antipolis), Rustrel-Pays d'Apt, France for their many significant contributions to this measurement. FG was supported by grant SFRH/BPD/13995/2003 of the Portuguese Foundation for Science and Technology (FCT); TM, by CFN-275-BPD-01/06 of the Nuclear Physics Center of the University of Lisbon. This work was supported in part by POCI grants FP/63407/2005 and FIS/57834/2004 of the FCT, co-financed by FEDER.